\documentclass[a4paper,10pt]{article}
\usepackage[utf8]{inputenc}
\usepackage{amsmath}
\usepackage{anysize}
\usepackage{ latexsym}
\usepackage{ amssymb}
\usepackage{amsthm}
\usepackage{mathrsfs}
\usepackage{enumerate}
\usepackage{amsmath}
\usepackage{amsfonts}

\usepackage{booktabs}
 \usepackage{multirow}
\usepackage{mathrsfs}
\usepackage{amsfonts,latexsym}
\usepackage{amsmath, amsthm}

\usepackage[dvips]{graphicx}
\usepackage{amscd}
\usepackage{subfigure}
\usepackage{color}
\usepackage{algorithm}
\usepackage{algorithmic}
\usepackage{amssymb}
\usepackage{a4wide}
\usepackage{subfig}

\providecommand{\keywords}[1]{\textbf{{Key words:}} #1}

\usepackage[utf8]{inputenc}

\title{Solitary wave shoaling and breaking in a regularized Boussinesq system}
\author{ Amutha Senthilkumar
 \\ 
{\small Department of Mathematics, University of Bergen} \\
{\small Postbox  7803, 5020 Bergen, Norway} \\
{\small Email:  amutha.senthilkumar@math.uib.no }\\  
{\small Telephone number:  (47) 55584858}\\ \\
}

\begin{document}

\maketitle

\begin{abstract}
A coupled BBM system of equations is studied in the situation of   water waves propagating  over decreasing fluid depth. A conservation equation for mass and a wave breaking 
criterion  valid in the Boussinesq approximation is found. 
A Fourier collocation  method  coupled with a 4-stage Runge-Kutta time integration scheme is employed  to approximate solutions of the BBM system.  
The  mass  conservation equation is  used to quantify the role of reflection in the shoaling of solitary waves on a sloping bottom. 
Shoaling results based on an adiabatic approximation are analyzed. 
Wave shoaling and  the criterion of breaking  solitary waves on a  sloping bottom is studied. To validate the numerical model the simulation results are compared with those obtained by
Grilli et al. \cite{GSS1997}  and a good agreement between them is observed.
Shoaling of solitary waves of two different types of mild slope model systems in \cite{C2003} and \cite{MDE2009}  are compared, 
and  it is found that each of these models works well in their respective 
regimes of applicability.

\end{abstract}

\keywords{Coupled BBM system, Shoaling rates, Mass conservation law}

\section{Introduction}

Model equations for free surface water waves propagating in a 
horizontal channel of uniform depth have been widely studied  for many years.
Boussinesq models  incorporate  the lowest-order effects of nonlinearity
and frequency dispersion as corrections to the linear long wave equation. These models are widely used for describing the
propagation of non-linear shallow water waves near coastal regions.
In Boussinesq theory, it is important to  assume that water is incompressible, inviscid and the flow is irrotational.
There are two important parameters which  are the nonlinearity, the ratio of amplitude to depth,
represented by $\alpha=a / h_0$, and the dispersion, the ratio of depth to
wavelength, represented by $\beta=h_0^2/l^2$. As  explained in detail in \cite{BCS2002}, the Boussinesq
approximation is valid only when both $\alpha$ and $\beta$  
are small and have the same order of magnitude.
 
The  more realistic situation of uneven bottom profile
is  fundamental  to studies of
ocean wave dynamics in coastal regions. Several authors 
\cite{GN1976,   JO1973,  KCKD2000,  M1979,   TW1994, Wh1974}
have included the effect of smooth and slowly varying bottom topographies in both Boussinesq and shallow water theory. 
The `classical' Boussinesq model was applied to shallow water of
uneven bottom in two horizontal dimensions by Peregrine \cite{P1967},
who used depth-averaged velocity as a dependent variable and derived the system 
\begin{subequations} \label{eq:per}
 \begin{eqnarray} 
 \eta _t+\nabla \cdot \left [ (h+ \eta )\bar{\textbf{u}} \right ]&=&0,  \label{eq:per1}\\
\bar{\textbf{u}}_t+\nabla \eta +(\bar{\textbf{\textbf{u}}}\cdot \nabla )\bar{\textbf{u}}
- \frac{h}{2}\nabla (\nabla \cdot(h\bar{\textbf{u}}_t))
+ \frac{h^2}{6}\nabla (\nabla \cdot (\bar{\textbf{u}}_t))&=&0, \label{eq:per2}
\end{eqnarray}
\end{subequations}
where the independent variable $\textbf{x}=(x,y)$ represents the position, $\eta=\eta(\textbf{x},t)$ represents the 
deviation of the free surface from its rest position  at time $t$,  $\textbf{u}=\textbf{u}(\textbf{x},z,t)$ denotes the horizontal
velocity of the fluid at some height, while $\bar{\textbf{u}}$ denotes the depth-averaged velocity 
\begin{equation}\label{eq:uper}
 \bar{\textbf{u}}=\frac{1}{h+ \eta } \int_{-h }^{ \eta} {\textbf{u}} \,dz,
\end{equation}
 and the bottom is $z=-h(\textbf{x})$.

Several improved Boussinesq-type models have been developed,
starting with  Madsen et al. \cite{MMS1991},   Nwogu \cite{N1993} and Wei et al. \cite{WKGS1995}, among others.
Madsen et al. \cite{MMS1991} achieved an improved linearized model by rearranging higher-order terms in the  classical  momentum equations, 
which are formally equivalent to zero within the accuracy of the model.
Nwogu \cite{N1993} demonstrated the flexibility obtained by using the velocity at an arbitrary distance 
from the still water level as the velocity variable. Wei et al.  \cite{WKGS1995} used Nwogu's approach to derive a fully nonlinear  extensions of Boussinesq
equations which further extended the range of validity of Boussinesq models without the weak nonlinearity restriction.
 It is worth mentioning that in \cite{C2003, MDE2009} the Boussinesq model
\eqref{eq:per}  has been extended
to moving bottom topography, where the bottom topography depends on $x$, $y$ and $t$. 
In the paper \cite{MDE2009},  a BBM-type system  (see \cite{BBM}) has been derived and solved numerically  using  a finite element method.
One way in which the BBM system differs from Peregrine's Boussinesq system
is in the way it is amenable to numerical integration. Indeed, it is much easier to
define a stable numerical approximation to a system of BBM type than to other
Boussinesq systems, such as the Peregrine system.
On the other hand, the Peregrine system features exact mass conservation while
mass conservation in the BBM-type systems is only approximate.
Nevertheless, in the current work, we use a system of BBM type for numerical convenience.

The main contribution of the present paper is an in-depth study of wave  reflection in a shoaling analysis based on 
of Boussinesq systems such as \eqref{eq:per}. Wave shoaling is the effect by which surface waves  propagating shorewards experience a decrease in the water depth. The study of  shoaling waves is of importance in the nearshore 
areas, and in the design of coastal structures. As part of our analysis, we formulate an
approximate mass balance law associated to the Boussinesq scaling, such as developed for flat bottoms
in \cite{AK2012}. We also extend the wave breaking criterion from \cite{BK2011} to the case of uneven beds.
The mass balance equation is used in quantifying wave reflection due to the bottom slope, 
and the wave breaking criterion is  used to determine an approximate termination point for the shoaling curves.
 A significant amount of literature has focused on the use of nonlinear shallow water equations to analyse long wave 
shoaling on a mildly sloping beach, and both experimental and numerical investigations have been carried out.
However, reflection has not been quantified.

Many experimental studies, including the early studies    \cite{CS1969, IK1954} were aimed partly at comparison with classical shoaling laws such as the laws of Green and
Boussinesq. However, most experimental work on wave shoaling has shown that actual shoaling curves vary
considerably from the predictions of both Green’s and Boussinesq’s law. 
Grilli et al.\cite{GSS1997} solved the full Euler equations by direct
numerical integration and this work compares their shoaling results with the numerical solution obtained from the present work.

Wave breaking is also important in studying nearshore area phenomena  
and is also important for the study of tsunami propagation in coastal regions, 
because solitary waves are often used to model steep surface waves shoaling on beaches. 
An enormous literature also exists about  breaking waves in a number of situations, including shoaling,
wave breaking in open bodies of water, and breaking induced by a
wavemaker (see \cite{D2001, Sv2006}, for instance). Chou and Ouyang  \cite{CQa1999, CQb1999} and 
Chou et al.  \cite{CSY2003} discussed the criterion for the breaking of solitary waves on different slopes using the boundary element method to simulate the process of wave breaking. 
Using the fully nonlinear potential flow wave model,  Grilli et al.  \cite{GSS1997} derived a criterion for  wave breaking.
In this paper, a different criterion of breaking solitary waves on a sloping bottom of  BBM-type system is derived based on previous work in \cite{BK2011}.  
Characteristics such as the breaking index, the waveheight, the water depth and the maximum particle velocity 
at the breaking point are studied  and the breaking indices  are compared with those obtained by
Grilli et al. \cite{GSS1997} and Chou et al. \cite{CSY2003}. 

The present paper is organized as follows. In Section \ref{sec:Derivation}, the outline for  
the derivation of the coupled BBM-type system \cite{MDE2009} is given,  and 
also the mass balance equations and the wave breaking criterion are derived. 
In Section \ref{sec:numericalmethods}, the  coupled BBM-type system is solved numerically using a Fourier collocation  method  coupled with a 4-stage Runge-Kutta time integration scheme.
We  validate  the convergence of the numerical scheme and demonstrate the effectiveness of the numerical method applied to 
our model system in simulations of solitary wave shoaling  on a sloping bottom.
Mass reflection, shoaling and wave breaking are studied numerically. This paper  compares  two  models;    
the coupled BBM-type system derived by Chen \cite{C2003} and  the one in Mitsotakis \cite{MDE2009}
with respect to evolution of solitary waves. This comparison  is concerned with initial  wave profiles and wave shoaling on  slopes that correspond to unidirectional propagation.
Finally, a short conclusion is given in Section \ref{sec:conclusion}.

\section{Derivation of the system}
\label{sec:Derivation}
The main model system to be used here belongs to a family of models derived in Mitsotakis \cite{MDE2009}. Let us briefly outlined the derivation.
In order to obtain the Boussinesq system,  the full water wave problem is used. A Cartesian coordinate system $(x,z)$ is considered,
with the $x$- axis along the still water level and $z$- axis pointing vertically upwards. The fluid domain is bounded by the sea bed at $z=-h(x)$  and the free surface  $z=\eta(x,t)$. 
Then  the  system of Euler equations 
for potential flow theory in the presence of a free surface is used.
The derivation of the Boussinesq system is only briefly sketched.
For a full derivation, the interested reader may consult \cite{C2003} and \cite{MDE2009}.
The variables are non-dimensionalized using following scaling:
\begin{subequations} \label{eq:nondimensional}
\begin{eqnarray}
 \tilde{x}=\frac{x}{l},  \ \ \ \tilde{z}=\frac{z}{h_0},
 \ \ \ \tilde{t}=\frac{\sqrt{gh_0}t}{l},\\
\text{ and } \ \ 
\tilde{h}=\frac{h}{h_0}, \ \ \ \ \tilde{\eta}=\frac{\eta}{a}, \ \ \ \tilde{\phi}=\frac{h_0}{al\sqrt{gh_0}}\phi,
\end{eqnarray}
\end{subequations}
where tilde ($\tilde{\quad}$) denotes non-dimensional variables, and $h_0$, $l$ and $a$ denote  characteristic water depth,    wavelength and   wave amplitude, respectively. 

Consider a standard asymptotic expansion of the velocity potential $\phi$ and using the Laplace  condition ($ \triangle \phi=0, \  -h<z<\eta$),  write the velocity potential $\tilde{\phi}$ in the simplest form
\begin{equation}\label{eq:phi}
 \tilde{\phi}= \tilde{\phi}^{(0)}+\frac{\tilde{z}}{1!}\tilde{\phi}^{(1)}+(-\beta )\left [ \frac{\tilde{z}^2}{2!}\frac{\partial ^ 2 }{\partial \tilde{x}^2}\tilde{\phi}^{(0)}
 +\frac{\tilde{z}^3}{3!}\frac{\partial ^2 }{\partial \tilde{x}^2}\tilde{\phi}^{(1)}\right ]+(\beta ^2 )\left [ \frac{\tilde{z}^4}{4!}\frac{\partial^4}{\partial \tilde{x}^4}\tilde{\phi}^{(0)}
 +\frac{\tilde{z}^5}{5!}\frac{\partial ^ 4 }{\partial \tilde{x}^4}\tilde{\phi}^{(1)}\right ]+\mathcal{O}(\beta ^3),
\end{equation}
which is a series solution with only two unknown functions $\tilde{\phi}^{(0)}$ and $\tilde{\phi}^{(1)}$.  Then the velocity field can be expressed as
\begin{subequations} \label{eq:velocities}
 \begin{eqnarray}
  \tilde{u}(\tilde{x},\tilde{z},\tilde{t})&=&\tilde{\phi}_{\tilde{x}}=
\hat{u}+\beta \left [ \frac{\tilde{z}}{1!} \hat{w}_{\tilde{x}}- \frac{\tilde{z}^2}{2!}\hat{u}_{\tilde{x}\tilde{x}}\right ]
+ \beta ^2 \left [ -\frac{\tilde{z}^3}{3!} \hat{w}_{\tilde{x}\tilde{x}\tilde{x}}+\frac{\tilde{z}^4}{4!}\hat{u}_{\tilde{x}\tilde{x}\tilde{x}\tilde{x}}\right ]+\mathcal{O}(\beta ^3),\label{eq:velocities1} \\ 
\tilde{w}(\tilde{x},\tilde{z},\tilde{t})&=&\tilde{\phi}_{\tilde{z}}=
\beta \left [  \hat{w}-\tilde{z} \hat{u}_{\tilde{x}}\right ]+ \beta ^2 \left [ -\frac{\tilde{z}^2}{2!} \hat{w}_{\tilde{x}\tilde{x}}
+\frac{\tilde{z}^3}{3!}\hat{u}_{\tilde{x}\tilde{x}\tilde{x}}\right ]+\mathcal{O}(\beta ^3), \label{eq:velocities2}
 \end{eqnarray}
\end{subequations}
where $\hat{u}$ and $\hat{w}$ are the velocities at $\tilde{z}=0$ and given by $\hat{u}=\tilde{\phi}_{\tilde{x}}^{(0)},\ \hat{w}=(1/ \beta) \tilde{\phi}^{(1)}.$

In order to establish the relation between $\hat{u}$ and $\hat{w}$,  use the bottom kinematic boundary condition ($\phi_z+ h_x  \phi _x =0 \text{ at} \ z=-h$), which has the following form after 
substituting  the above asymptotic expressions: 
\begin{equation}\label{eq:wi}
 \hat{w}=-(\tilde{h} \hat{u})_{\tilde{x}}+\beta 
\frac{\partial }{\partial \tilde{x}} \left ( \frac{\tilde{h}^3}{3!} \hat{u}_{\tilde{x}\tilde{x}} - \frac{\tilde{h}^2}{2!} (\tilde{h}\hat{u})_{\tilde{x}\tilde{x}}\right )+\mathcal{O}(\beta ^2).
\end{equation}
Now inserting (\ref{eq:phi}), (\ref{eq:velocities}) and (\ref{eq:wi})  into free surface boundary conditions, one may derive the following Boussinesq system with variable bottom
\begin{subequations} \label{eq:system}
 \begin{eqnarray}
  \hat{u}_{\tilde{t}}+{\tilde{\eta }}_{\tilde{x}}+\alpha \hat{u}\hat{u}_{\tilde{x}}&=& \mathcal{O}(\alpha \beta,\beta ^2), \label{eq:system1} \\
{\tilde{\eta }}_{\tilde{t }}+\left ( \alpha {\tilde{\eta }}\hat{u}+\tilde{h}\hat{u} \right )_{\tilde{x}}-\beta 
\frac{\partial }{\partial \tilde{x}} \left ( \frac{\tilde{h}^3}{3!} \hat{u}_{\tilde{x}\tilde{x}}
- \frac{\tilde{h}^2}{2!} (\tilde{h}\hat{u})_{\tilde{x}\tilde{x}}\right )&=&\mathcal{O}(\alpha \beta,\beta ^2). \label{eq:system2}
 \end{eqnarray}
\end{subequations}
It is emphasized that from the above system, and  in terms of $\hat{u}$, one can extend the system in terms of other velocity variables, such as the velocity at an arbitrary $z$ location. In this work we use a trick due to \cite{N1993}.
Namely,   a new velocity variable $\tilde{u} ^ {\theta}$ defined  at an arbitrary water level $\tilde{z} =-\tilde{h}+\theta(\alpha \tilde{\eta}+ \tilde{h}) $, 
with $0 \leq \theta \leq 1 $. Applying  the standard techniques of inversion it is not difficult to  derive the following  expression as an asymptotic formula for   $\hat{u}$ in terms of $\tilde{u}^{\theta }$: 
\begin{equation}\label{eq:velocitiesu}
\hat{u}=\tilde{u}^{\theta }+\beta \left ( \tilde{h}(\theta-1)(\tilde{h}\tilde{u}^{\theta })_{\tilde{x}\tilde{x}} 
+(\tilde{h})^2(\theta-1)^2\frac{1}{2!} (\tilde{u}^{\theta })_{\tilde{x}\tilde{x}}\right )+\mathcal{O}(\alpha \beta,\beta ^2).
 \end{equation}
Switching to the variable  $\tilde{u}^{\theta }$,  the following expressions are obtained:
\begin{equation}\label{eq:linearsystem}
 {\tilde{\eta }}_{\tilde{t }}=-\left ( \tilde{h}\tilde{u}^{\theta } \right )_{\tilde{x}}+\mathcal{O}(\alpha, \beta),  \ \ \ \  \
 \tilde{u}^{\theta }_{\tilde{t }}=-{\tilde{\eta }}_{\tilde{x }}+\mathcal{O}(\alpha, \beta).
\end{equation}
Following the methodology in \cite{BCS2002}, for arbitrary $\mu, \nu \in \mathbb{R} $ and using (\ref{eq:linearsystem}),  the following equations are derived
\begin{subequations} \label{eq:muanu}
 \begin{eqnarray}
  ( \tilde{h}\tilde{u}^{\theta } )_{\tilde{x}\tilde{x}}=\mu ( \tilde{h}\tilde{u}^{\theta } )_{\tilde{x}\tilde{x}}-(1-\mu ){\tilde{\eta }}_{\tilde{t }\tilde{x}}+\mathcal{O}(\alpha, \beta) \\
\tilde{u}^{\theta }_{\tilde{t }\tilde{x}\tilde{x}}=(1-\nu ) \tilde{u}^{\theta }_{\tilde{t }\tilde{x}\tilde{x}}-\nu{\tilde{\eta }}_{\tilde{x}\tilde{x}\tilde{x}}+\mathcal{O}(\alpha, \beta) 
 \end{eqnarray}
\end{subequations}
Using equations  (\ref{eq:system})-(\ref{eq:muanu}) and  appropriate expansions, the following system is derived:

\begin{subequations} \label{eq:abcdsystem}
 \begin{eqnarray}
  \tilde{u}^{\theta }_{\tilde{t}}+{\tilde{\eta }}_{\tilde{x}}+\alpha \tilde{u}^{\theta }\tilde{u}^{\theta }_{\tilde{x}}+\beta  \left \{ B \tilde{h}  \left [(\tilde{h}_{\tilde{x}}\tilde{\eta}_{\tilde{x}})_ {\tilde{x}}+\tilde{h}_{\tilde{x}}\tilde{\eta}_{\tilde{x}\tilde{x}}
\right ] +c \tilde{h}^2\tilde{\eta}_{\tilde{x} \tilde{x} \tilde{x}} -d \tilde{h}^2\tilde{u}^{\theta }_{\tilde{x}\tilde{x}\tilde{t}} \right \} =\mathcal{O}(\alpha \beta,\beta ^2)  \\
{\tilde{\eta }}_{\tilde{t }}+\left ( \alpha {\tilde{\eta }}\tilde{u}^{\theta }+\tilde{h}\tilde{u}^{\theta } \right )_{\tilde{x}}
+\beta \frac{\partial }{\partial \tilde{x}}\left \{   A \tilde{h}^2 \left [ (\tilde{h}_ {\tilde{x}}\tilde{u}^{\theta })_{\tilde{x}}+\tilde{h}_ {\tilde{x}}\tilde{u}^{\theta }_ {\tilde{x}}\right ] +a{\tilde{h}^2}(\tilde{h}\tilde{u}^{\theta })_ {\tilde{x}\tilde{x}}
 -b \tilde{h}^2 \tilde{\eta}_{\tilde{x}\tilde{t}} \right \}=\mathcal{O}(\alpha \beta,\beta ^2).
  \end{eqnarray}
\end{subequations}
The parameters a, b, c and d are the same as in \cite{BCS2002}, where
\begin{eqnarray}
A=\frac{1}{2}\left [ \frac{1}{3}-(\theta -1)^2 \right ], \ \ && B=1-\theta, \nonumber\\
a=\frac{1}{2}\left (  \theta ^2-\frac{1}{3}\right ) \mu, \ \ && b=\frac{1}{2}\left (  \theta ^2-\frac{1}{3}\right ) (1-\mu), \nonumber\\
c=\frac{1}{2}\left ( 1- \theta ^2\right ) \nu,\ \ && d=\frac{1}{2}\left ( 1- \theta ^2\right ) (1-\nu). 
\end{eqnarray}

Note that the coupled BBM-type system appears in (\ref{eq:abcdsystem}) if $\mu=0$ and $\nu=0$. Disregarding terms of order $\mathcal{O}(\alpha \beta,\beta ^2)$ and dropping the superscript $\theta$, the system  takes the following form in dimensional variables
\begin{subequations} \label{eq:BBmsystem}
 \begin{eqnarray}
u_{t}+g {\eta }_{x}+ {u}{u}_{x}+  2 B g{h}  {h}_{x}{\eta}_{xx}+B g{h}  {h}_{xx}{\eta}_{x}
 -d {h}^2 u_{xxt}  =0, \\
{\eta }_{t}+\left ( {{\eta }}u+{h}{u} \right )_{x}
+ \frac{\partial }{\partial x}\left \{  2A{h}^2 h_ {x}u_ {x}+A{h}^2 h_ {xx}u
 - b{h}^2 {\eta}_{xt} \right \}=0.
 \end{eqnarray}
\end{subequations}
Assuming  the depth $h$ is constant,  the above system reduces to the original coupled BBM system  in \cite{BCS2002}.

\subsection{Mass balance}
\label{subsec:massbalance}
As mentioned in the introduction, the use of the BBM system necessitates the derivation of an approximate mass balance law. 
The following mass balance derivation is based on the work in \cite{AK2012}. They have already presented mass balance theory for the Boussinesq models with even bottom profile. Since we are interested in varying bottom topography,
we provide the following derivation. The integral form of the equation of mass conservation is
\begin{equation}
 \frac{d}{dt}\int_{x_1}^{x_2}\int_{-h(x)}^{\eta} \rho \,dz \,dx =\left [ \int_{-h(x)}^{\eta} \rho \phi _x \,dz \right ]_{x_2}^{x_1},
\end{equation}
since there is no mass flux through the bottom or through the free surface. In non-dimensional variables the above relation becomes
\begin{equation}
 \frac{d}{d\tilde{t}}\int_{\tilde{x}_1}^{\tilde{x}_2}\int_{-\tilde{h}}^{\alpha \tilde{\eta}} \,d \tilde{z} \,d\tilde{x} =\left [ \int_{-\tilde{h}}^{\alpha \tilde{\eta}}  \alpha \tilde{\phi}_{\tilde{x}}\,d \tilde{z} \right ]_{\tilde{x}_2}^{\tilde{x}_1}.
\end{equation}
After integration with respect to $ \tilde{z}$ and use of asymptotic expansion of $\tilde{\phi}$, we obtain
\begin{equation}
 \int_{\tilde{x}_1}^{\tilde{x}_2}(\alpha \tilde{\eta}+\tilde{h})_{\tilde{t}}  \,d\tilde{x} =\alpha \left [ \hat{u}( \tilde{h}+\alpha \tilde{\eta })
 +\frac{\tilde{h}^2}{2!}\beta (\hat{u} \tilde{h})_{\tilde{x}\tilde{x}}-\frac{\tilde{h}^3}{3!}\beta (\hat{u} )_{\tilde{x}\tilde{x}}\right ]_{\tilde{x}_2}^{\tilde{x}_1}+\mathcal{O}(\alpha \beta,\beta ^2).
\end{equation}
Note that if we take the limit  $\tilde{x}_2 \to \tilde{x}_1$, where $\tilde{x}_2 =x_2/l \text{ and }   \tilde{x}_1=x_1/l$,   then we obtain the balance equation (\ref{eq:system2}).
i.e, 
\begin{equation} \label{eq:massbalance}
 \frac{\partial } {\partial \tilde{t}}\tilde{M}+ \frac{\partial } {\partial \tilde{x}}\tilde{q}_M=\mathcal{O}(\alpha \beta,\beta ^2),
\end{equation}
where 
\begin{equation*}
 \tilde{M}=\alpha \tilde{\eta }+\tilde{h}, \ \  \ \ \tilde{q}_M= \alpha \left [  \left ( \alpha {\tilde{\eta }}\tilde{u}^{\theta }+\tilde{h}\tilde{u}^{\theta } \right )
+\beta (\theta -\textstyle\frac{1}{2}) \tilde{h}^2 ( \tilde{h}\tilde{u}^{\theta })_{\tilde{x}\tilde{x}}
+\beta {\tilde{h}^3}(\textstyle\frac{1}{2}(\theta -1)^2-\textstyle\frac{1}{6}) (\tilde{u}^{\theta })_{\tilde{x}\tilde{x}}\right ].
\end{equation*}
The derivation could also be based on the differential form of the mass conservation, such as in \cite{AK2014}.
If we use the scalings $M=\rho h_0 \tilde{M}$ and $q_M=\rho h_o \sqrt{(gh_0)} \tilde{q}_M$, then the dimensional form of these quantities are
\begin{equation}\label{eq:mass}
 {M}=\rho (\eta+h(x)), \ \  \ \ {q}_M= \rho  \left [  u(h+\eta)+h^2(\theta-\textstyle \frac{1}{2})(hu)_{xx}+\textstyle \frac{1}{2}h^3((\theta-1)^2-\textstyle \frac{1}{3})u_{xx}\right ].
\end{equation}
Eq. (\ref{eq:massbalance}) represents the approximate mass balance equation. The net mass transfer to or from a control volume during a time interval $\bigtriangleup t$ is equal to the net change (increase or decrease) in the total mass in the control volume 
during $\bigtriangleup t$.  
 In \cite{AK2012}, they proved that the maximum error in the conservation of mass is smaller than   $\mathcal{O}(\alpha \beta,\beta ^2)$ in the case of even bottom profile using a coupled BBM system.
 In Subsection (\ref{subsec:massconservation}) the amount of mass reflection will be computed for different cases.

\subsection{Wave breaking in BBM model system}
\label{subsec:breakcriteria}
As waves approach the shoreline the wavelength and phase velocity decrease and the wave amplitude grows larger. The wave then crashes onto shore  because it becomes too steep for the bottom of the wave to carry it. 
The breaking of waves mostly depends on  wave steepness and beach slope. As explained in \cite{BK2011},  if the horizontal velocity  near the crest of a wave exceeds the celerity of the wave, then the wave breaks. 
Let us denote propagation speed  by $U$ and horizontal velocity by u. 
The horizontal velocity $u$  can be obtained from  (\ref{eq:velocities1}) and (\ref{eq:velocitiesu}):
\begin{equation}\label{eq:velocitiesuu}
\tilde{u}=\tilde{u}^{\theta }+\beta \left ( (\tilde{h}(\theta-1)-\tilde{z})(\tilde{h}\tilde{u}^{\theta })_{\tilde{x}\tilde{x}} 
+((\tilde{h})^2(\theta-1)^2-\tilde{z}^2)\frac{1}{2!} (\tilde{u}^{\theta })_{\tilde{x}\tilde{x}}\right )+\mathcal{O}(\alpha \beta,\beta ^2).
 \end{equation}
 It is evident that once $u^\theta (x,t)$ is known,  (\ref{eq:velocitiesuu}) can be used to
approximate the horizontal velocity at any depth.
After neglecting the second-order term, the dimensional form of the equation is given by
\begin{equation}\label{eq:velocitiesuuu}
{u}={u}^{\theta }+ ({h}(\theta-1)-{z})({h}{u}^{\theta })_{{x}{x}} 
+({h}^2(\theta-1)^2-{z}^2)\frac{1}{2!} ({u}^{\theta })_{\tilde{x}\tilde{x}}.
 \end{equation}
Wave breaking occurs if
\begin{equation}\label{eq:bvelocities}
{u}^{\theta }+ ({h}(\theta-1)-{\eta})({h}{u}^{\theta })_{{x}{x}} 
+({h}^2(\theta-1)^2-{\eta}^2)\frac{1}{2!} ({u}^{\theta })_{{x}{x}} > U.
\end{equation}
Since the fluid domain depends on the surface profile, the value $z=\eta $ is used to approximate velocities near the surface.
It is clear that the solutions  $\eta(x,t)$ and $u^{\theta }(x,t)$ of the system (\ref{eq:BBmsystem}) and propagation speed   $U$ are needed to find the breaking criterion.

\section{Numerical methods}
\label{sec:numericalmethods}
The system (\ref{eq:BBmsystem}) has been solved numerically using a Fourier collocation  method  coupled with a 4-stage Runge-Kutta time integration scheme.
For numerical computations,  periodic boundary conditions on the domain $ [0,L]$ are used. For this
  the problem is translated to the interval $[0, 2 \pi ]$ using the scaling $u(\lambda x,t)=v(x, t),\ \  \eta(\lambda x,t)=\xi(x, t) \text{ and } h(\lambda x)=h_1(x), $ where $\lambda=\textstyle \frac{L}{2 \pi}$.
Then the BBM- system (\ref{eq:BBmsystem}) becomes
\begin{subequations} \label{eq:lBBmsystem}
 \begin{eqnarray*}
\lambda^3 v_{t}+\lambda^2 g {\xi}_{x}+\lambda^2 {v}{v}_{x}+  2 B g{h_1}  {h_1}_{x}{\xi}_{xx}+ B g{h_1}  {h_1}_{xx}{\xi}_{x}
 -\lambda d{h_1}^2 v_{xxt}  &=&0, \ \ x \in [0,2\pi], \\
\lambda^3 {\xi }_{t}+\lambda^2 \left ( {{\xi}}v+{h_1}{v} \right )_{x}
+ \frac{\partial }{\partial x}\left \{  2A{h_1}^2 h_ {1x}v_ {x}+A{h_1}^2h_{1xx}v
 - \lambda b{h_1}^2 {\xi}_{xt} \right \}&=&0, \ \ x \in [0,2\pi], \\
v(x,0) =u(\lambda x,0), \ \ \ \xi(x,0) =\eta(\lambda x,0), \\  
v(0,t)=v(2\pi, t), \ \ \  \xi(0,t)=\xi(2\pi, t), \text{  for } t \geq 0.
 \end{eqnarray*}
\end{subequations}
 Consider the set of $ N$ evenly spaced grid points $x_j=\textstyle \frac{2 \pi j}{N},\ \ j =1,...N$ in the interval $[0, 2 \pi ]$ referred to as collocation nodes. 
 The spectral-collocation method is implemented in the physical space by 
 seeking approximate solutions through a global periodic interpolation polynomial  of the form
 \begin{equation*}
 v_N (x)=  \sum_{j=1}^{N}v_N(x_j)g_j(x), \ \  \xi_N (x)=  \sum_{j=1}^{N} \xi_N(x_j)g_j(x),
\end{equation*}
where $g_j(x)=\frac{1}{N} \sin \left ( \frac{N(x-x_j)}{2} \right ) \cot \left ( \frac{1}{2}(x-x_j) \right )$ 
and $v_N(x), \ \xi_N(x)$ is an interpolation of the function $v(x), \ \xi(x)$ respectively, i.e., $v_N(x_j)=v(x_j), \xi_N(x_j)=\xi(x_j)$ (see \cite{T2000}, \cite{GO1986}). Moreover, the corresponding Fourier collocation 
differentiation matrices 
$D_x$ and $D_{xx}$ are given by
 \begin{subequations} \label{eq:derivatives}
 \begin{eqnarray}
  D_{ij}^{(1)}= \frac{dg_j}{dx} (x_i)=\left\{\begin{matrix}
\frac{1}{2}(-1)^j \cot (\frac{x_i-x_j}{2}) & i \neq j   \\ 
 0 &  i=j
\end{matrix}\right. \label{eq:derivatives1} \\
D_{ij}^{(2)}=\frac{d^2g_j}{dx^2} (x_i)=\left\{\begin{matrix}
-\frac{(-1)^j}{2\sin ^2 ((x_i-x_j)/2)}  & i \neq j  \\ 
 \frac{- \pi^2 }{3h^2}-\frac{1}{6} &  i = j \label{eq:derivatives2}
\end{matrix}\right. 
 \end{eqnarray}
\end{subequations}
 Then at the collocation points $x=xj$, the system becomes
 \begin{subequations}
 \begin{eqnarray*}
  \left [ \lambda ^3 I_N- \lambda b \ D_N diag(h_1^2)D_N \right ]\xi_{Nt}&=& -\lambda^2 D_N(diag(h_1)v_N)-\lambda^2 D_N(\xi_N v_N)\\
  \:&&-D_N(2Ah_{1} ^2 h_{1x}  D_N(v_N)+ Ah_{1xx} h_1^2  v_N), \\
 \left [ \lambda ^3 I_N-\lambda d  \ diag(h_1^2)D_N^{(2)}\right ]v_{Nt}&=& -\lambda^2 g D_N(\xi_N)-\lambda^2 (0.5)D_N( v_N^2 ) \\
  \:&&-2Bg h_1 h_{1x}  D_N^{(2)}(\xi_N) -Bg h_1 h_{1xx}  D_N(\xi_N),
  \end{eqnarray*}
 \end{subequations}
where $I_N$ is the unit $N \times N$ matrix and $D_N$, $D_N^{(2)}$ are square matrices of dimensions $N \times N$ following from (\ref{eq:derivatives1}) and (\ref{eq:derivatives2}), respectively and $diag(h_1)$, $diag(h_1^2)$ are
the diagonal matrices of $h_1$ and $h_1^2$, respectively.
 This is a system of $N$ ordinary differential equations for $\xi_N$ and also $v_N$. The system is solved by using a fourth order explicit Runge-Kutta scheme with time step $\bigtriangleup t$.

 \subsection{Convergence study}
\label{subsec:convergence}
It is  important to verify  the convergence of the numerical scheme. This is done following  \cite{A2013}.  
A  numerical method is convergent if the numerically computed solution approaches the exact solution as the step size approaches 0. To test the convergence of these numerical methods,  the following discrete $L^2$ - norm is used
\begin{equation*}
\left \|\xi \right \|_{N,2}^2=\frac{1}{N}\sum_{j=1}^{N}\left | \xi(x_j) \right |^2,
\end{equation*}
and the  corresponding relative $L^2$ - error is then defined to be
\begin{equation*}
\frac{\left \| \xi -\xi_N\right \|_{N,2}}{\left \| \xi \right \|_{N,2}},
\end{equation*}
where $\xi_N(x_j)$ is the approximated numerical solution and $\xi(x_j)$ is the exact solution at a time T, for $j=1,2, \dots , N$. 

Supposing the case of an even bottom, the coupled BBM system features solitary-wave solutions in a closed form if $\theta ^2 =\textstyle \frac{7}{9}$ (see \cite{Ch1998}).
Since the analysis of the solitary wave shoaling and breaking given here depend on the exact formula for the solitary wave,  $\theta ^2 =\textstyle \frac{7}{9}$ is used in the present work.
Then the exact solitary wave solutions of system of equations (\ref{eq:BBmsystem}) takes the form
 
\begin{eqnarray} \label{eq:sbbm}
\eta(x,t)= H_0 \text{ sech}^2(\kappa_0 (x-C_0t)),\\
u(x,t) = W_0 \text{ sech}^2(\kappa_0 (x-C_0t)),
\end{eqnarray}
where $h_0$ is the undisturbed depth, $H_0$ is wave amplitude,
and the constants $W_0$, $C_0$ and $\kappa_0$ are given by
\begin{equation*}
W_0=\sqrt{\frac{3g}{H_0+ 3 h_0}}H_0, \ \ 
C_0=\frac{3h_0+2 H_0}{\sqrt{3h_0(H_0+3h_0)}}\sqrt{gh_0} \ \ 
\text{ and }
\kappa_0 =\frac{3}{2h_0}\sqrt{\frac{H_0}{2H_0 + 3h_0}}.
\end{equation*}
To check the convergence of these methods, we determine the $L^2$ - error each time for
n steps and set the step size  as  $ \bigtriangleup t =(t_{max}-t_{min})/ n $ for different n values   $n= 20,40,80,...$ 
( Table   \ref{tab:error time}) and different number of grid points $N= 256, 512, 1024,...$  ( Table   \ref{tab:error time1}) in the case of an even bottom topography. A representative result for  a wave of amplitude $0.5$ is 
given in Tables  \ref{tab:error time}  and  \ref{tab:error time1}. The numerical scheme was implemented in MATLAB. In this calculation, the solution was approximated from $T=0$ to $T=5$ and the size of the domain was L=100.
In the computations shown in  Table  \ref{tab:error time}, $N=1024$ Fourier modes were used.  Table   \ref{tab:error time} shows fourth-order convergence of the Runge–Kutta method in terms of the time step $\bigtriangleup t$. 
The 4th-order convergence of the scheme is apparent up to $\bigtriangleup t=0.0039$, when the error became dominated by the spatial discretization and the artificial periodicity.
Table  \ref{tab:error time1} shows the results of some computations aimed at validating the spatial convergence of the code. As expected, spectral convergence  
in terms of the number of spatial grid points $N$ is achieved in these computations. Computations were also performed for other solitary waves with heights between 0.1 and 0.6, and similar results were obtained for these cases.

\begin{table}[H]\footnotesize
  \begin{center}
    \begin{tabular}{| r | r | r | r | }
    \hline
    n \ \ \ &  $\bigtriangleup t \ \ \ $& $L^2$ - error  \ \ \ & Convergence rate \ \ \  \\ \hline
    
  \ \ \ 20 \ \ \ &  \ \ \ 0.2500 \ \ \ &  \ \ \  5.33e-02 \ \ \ & \ \ \   -  \ \ \ \\ \hline
 \ \ \ 40 \ \ \ & \ \ \ 0.1250 \ \ \ &  \ \ \ 3.93e-03  \ \ \ &   \ \ \  13.58 \ \ \     \\ \hline
  \ \ \ 80 \ \ \ & \ \ \ 0.0625 \ \ \ &  \ \ \   2.39e-04 \ \ \  &   \ \ \   16.44  \ \ \ \\ \hline
  \ \ \ 160 \ \ \ & \ \ \ 0.0312 \ \ \ &  \ \ \      1.44e-05 \ \ \ &   \ \ \  16.49   \ \ \   \\ \hline
 \ \ \  320 \ \ \ & \ \ \ 0.0156 \ \ \ &  \ \ \  8.89e-07 \ \ \ &   \ \ \   16.29  \ \ \  \\ \hline
  \ \ \ 640 \ \ \ & \ \ \ 0.0078 \ \ \ &  \ \ \  5.50e-08 \ \ \ &   \ \ \   16.15  \ \ \  \\ \hline
  \ \ \ 1280 \ \ \ & \ \ \ 0.0039 \ \ \ &  \ \ \  3.60e-09 \ \ \ &   \ \ \   15.35  \ \ \  \\ \hline
  \ \ \ 2560 \ \ \ & \ \ \ 0.0020 \ \ \ &  \ \ \  1.07e-09 \ \ \ &   \ \ \   03.36  \ \ \  \\ \hline
  \end{tabular}
\end{center}
 \caption{{$L^2$ - error  and convergence rate for Runge–Kutta method for different fixed step sizes in case of even bottom profile}}
\label{tab:error time}
\end{table}

\begin{table}[H]\footnotesize
  \begin{center}
    \begin{tabular}{| r | r | r | r | }
    \hline
   N \ \ \ &  $\bigtriangleup t$ \ \ \ & $L^2$ - error  \ \ \ & Convergence rate \ \ \ \\ \hline
    
  \ \ \ 256 \ \ \ &  \ \ \ 0.0001 \ \ \ & \ \ \  2.3 e-04 \ \ \ &  \ \ \  - \ \ \  \\ \hline
 \ \ \  512 \ \ \ & \ \ \ 0.0001 \ \ \ &  \ \ \  2.77e-09 \ \ \ &     \ \ \   84364.95   \ \ \    \\ \hline
  \ \ \ 1024 \ \ \  & \ \ \ 0.0001 \ \ \ & \ \ \ 3.09e-012  \ \ \ &   \ \ \    896.81 \ \ \  \\ \hline
  \ \ \ 2048 \ \ \  & \ \ \ 0.0001 \ \ \ &  \ \ \   5.371e-011 \ \ \  &   \ \ \  0.05  \ \ \   \\ \hline
  \end{tabular}
\end{center}
 \caption{$L^2$ - error  and convergence rate due to spatial discretization  in case of even bottom profile}
\label{tab:error time1}
\end{table}

 To Indicate the significance of the improvement,
Tables \ref{tab: Inhomogeneous error time} and  \ref{tab: Inhomogeneous error time1} show the results of computing approximate solutions of the inhomogeneous BBM-type system

\begin{subequations} \label{eq:Inhomogeneous BBmsystem}
 \begin{eqnarray}
u_{t}+g {\eta }_{x}+ {u}{u}_{x}+  2 B g{h}  {h}_{x}{\eta}_{xx}+B g{h}  {h}_{xx}{\eta}_{x}
 -d {h}^2 u_{xxt}  =f(x,t), \\
{\eta }_{t}+\left ( {{\eta }}u+{h}{u} \right )_{x}
+ \frac{\partial }{\partial x}\left \{  2A{h}^2 h_ {x}u_ {x}+A{h}^2 h_ {xx}u
 - b{h}^2 {\eta}_{xt} \right \}=g(x,t),
 \end{eqnarray}
\end{subequations}
where the functions $\eta(x,t)=0.3 \cos(x-t)$ and $ u(x,t)=0.3 \sin(x-t)$ are  used as the exact solutions and the bottom $h(x)=0.5-(0.1) \cos(x)$ is assumed.  
Then  the relative $L^2 -$ error for various pairs of combinations between the time step $\Delta t = 0.1/2n $, for $n = 1, 2, 3,...$; and $N = m \times 64$ for $m = 1, 2, 3,...$ is calculated. 
The results are shown in Table \ref{tab: Inhomogeneous error time} and Table \ref{tab: Inhomogeneous error time1}, where the solutions were approximated from $T = 0$ to $T = 5$.
These tables show that the  numerical implementation of BBM-type system with periodic  bottom function  $h(x)$ is correct. Similar results can be  obtained for other $2 \pi-$periodic functions $u$, $\eta$ and $h(x)$.

\begin{table}[H]\footnotesize
  \begin{center}
    \begin{tabular}{| r | r | r | r | }
    \hline
    n \ \ \ &  $\bigtriangleup t \ \ \ $& $L^2$ -error  \ \ \ & Convergence rate \ \ \  \\ \hline
          \ \ \ 50 \ \ \ &  \ \ \ 0.1000 \ \ \ &  \ \ \ 1.3046e-05 \ \ \ & \ \ \   -  \ \ \ \\ \hline
  \ \ \ 100 \ \ \ &  \ \ \ 0.0500 \ \ \ &  \ \ \ 8.2126e-07 \ \ \ & \ \ \  15.89  \ \ \ \\ \hline
  \ \ \ 200 \ \ \ & \ \ \ 0.0250 \ \ \ &  \ \ \ 5.1277e-08  \ \ \ &   \ \ \  16.02 \ \ \     \\ \hline
    \ \ \ 400 \ \ \ & \ \ \ 0.0125 \ \ \ &  \ \ \ 3.2023e-09  \ \ \  &   \ \ \ 16.01   \ \ \ \\ \hline
   \ \ \ 800 \ \ \ & \ \ \ 0.0063 \ \ \ &  \ \ \ 2.0007e-10  \ \ \  &   \ \ \ 16.01   \ \ \ \\ \hline
  \ \ \ 1600 \ \ \ & \ \ \ 0.0031 \ \ \ &  \ \ \ 1.2500e-11  \ \ \  &   \ \ \ 16.01   \ \ \ \\ \hline
   \ \ \ 3200 \ \ \ & \ \ \ 0.0016 \ \ \ &  \ \ \ 9.3000e-13  \ \ \  &   \ \ \ 13.41   \ \ \ \\ \hline
   \ \ \ 6400 \ \ \ & \ \ \ 0.0008 \ \ \ &  \ \ \ 6.2000e-13  \ \ \  &   \ \ \ 01.51   \ \ \ \\ \hline
  \end{tabular}
\end{center}
 \caption{Inhomogeneous BBM-type system (\ref{eq:Inhomogeneous BBmsystem}); $L^2$ - error  and convergence rate due to temporal discretization}
\label{tab: Inhomogeneous error time}
\end{table}

\begin{table}[H]\footnotesize
  \begin{center}
    \begin{tabular}{| r | r | r | r | }
    \hline
   N \ \ \ &  $\bigtriangleup t$ \ \ \ & $L^2$ -error  \ \ \ & Convergence rate  \ \ \ \\ \hline
      \ \ \ 64 \ \ \ &  \ \ \ 0.001 \ \ \ & \ \ \  9.6-01 \ \ \ &  \ \ \  - \ \ \  \\ \hline
     \ \ \ 128 \ \ \ &  \ \ \ 0.001 \ \ \ & \ \ \  5.4e-06 \ \ \ &  \ \ \  176234.99 \ \ \  \\ \hline
     \ \ \  256 \ \ \ & \ \ \ 0.001 \ \ \ &  \ \ \  2.1e-13 \ \ \ &     \ \ \  26437130.31  \ \ \    \\ \hline
  \ \ \ 512 \ \ \  & \ \ \ 0.001 \ \ \ & \ \ \ 6.6e-13  \ \ \ &   \ \ \    0.31 \ \ \  \\ \hline
  \ \ \ 1024 \ \ \  & \ \ \ 0.001 \ \ \ &  \ \ \  6.5e-13 \ \ \  &   \ \ \  1.007  \ \ \   \\ \hline
  \end{tabular}
\end{center}
 \caption{Inhomogeneous BBM-type system (\ref{eq:Inhomogeneous BBmsystem}); $L^2$ - error  and convergence rate due to spatial discretization  }
\label{tab: Inhomogeneous  error time1}
\end{table}

\subsection{Mass conservation results on a sloping bottom}
\label{subsec:massconservation}
The effect of  depth variations on solitary waves of  shallow water wave
theory is examined. The BBM system (\ref{eq:BBmsystem}) is simulated.  In all the numerical results of this subsection we use  $N = 1024,$  $\theta ^2 =\textstyle \frac{7}{9}$. 
Mass conservation is used to quantify the role of reflection in the shoaling of solitary waves. Note that the piecewise smooth linear bottom topography is used. To avoid the generation of small spurious oscillations  
due to the discontinuity in the derivative of the bottom function, it is smoothed near the singular points. 

 \begin{figure}[H]
\centering
\subfigure{
\includegraphics[width=.48\textwidth]{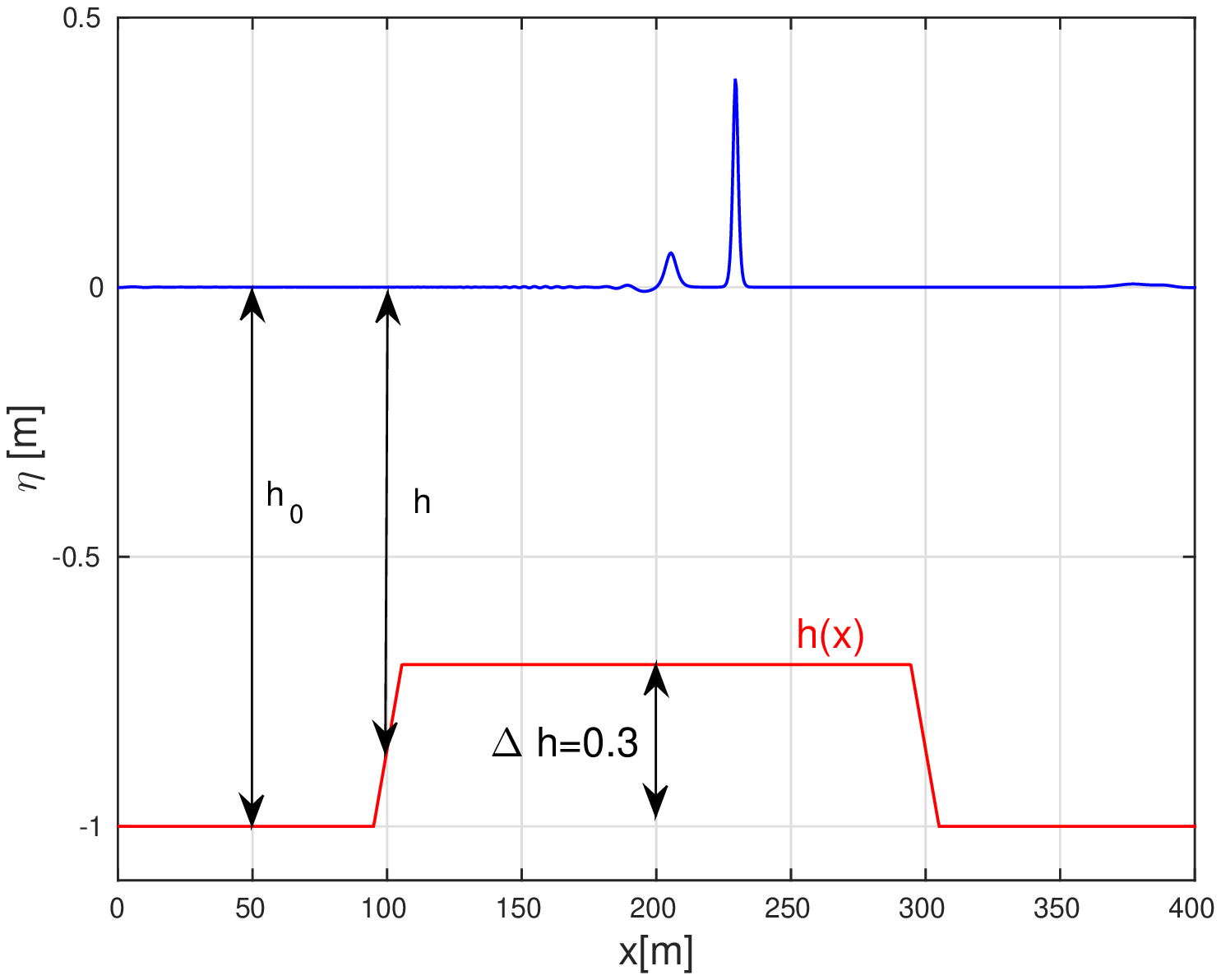}
}
\subfigure{
\includegraphics[width=.48\textwidth]{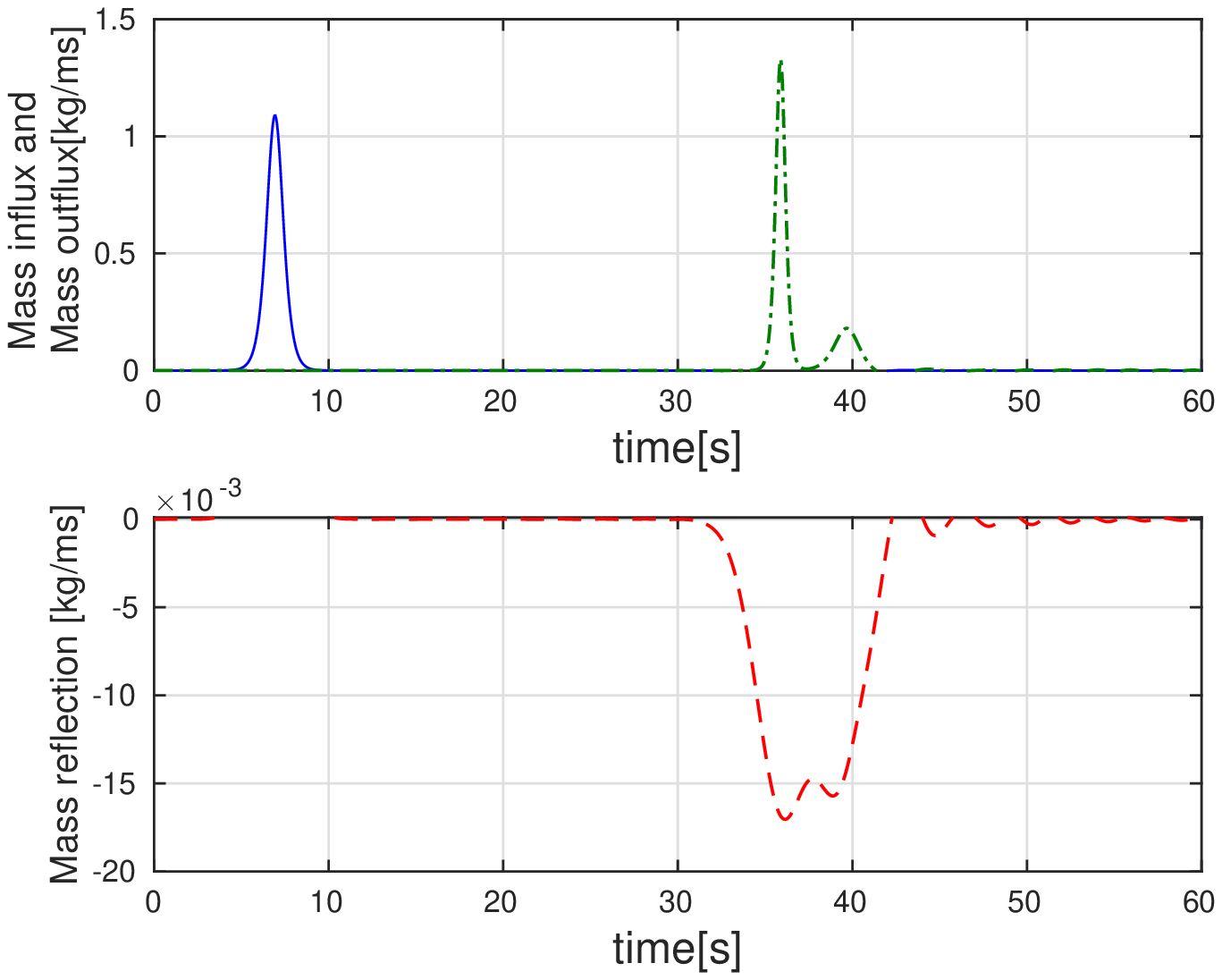}
}
\vskip-0.1in
\caption { The left panel shows a solitary wave solution for the system (\ref{eq:BBmsystem}) with initial amplitude 0.3m at time t=60s. 
The right panel  shows plots of time series of the mass influx at $x=50m$ (blue solid curve), the mass reflection at $x=50m$ (red dashed curve) and mass outflux at $x=150m$ (green dash-dotted  curve), per unit span.
The results are shown in the numerical domain.}
    \label{fig: massinfluxandmassoutflux}

\end{figure}

Consider a control volume delimited by the interval $[50,150]$ on the $x$-axis. The mass per unit width contained in this interval is defined by $\int_{50}^{150}M(x,t)\,dx$ 
and the mass flux through the boundaries of the control volume is defined by $q_M(50,t)$ and $q_M(150,t)$,  where $M$ and $q_M$ are given in (\ref{eq:mass}).
The quantities $M$ and $q_M$ during the passage of solitary wave are computed. It is  observed that the mass outflux is approximately equal to the addition of mass influx and the reflection of the mass.
In the right panel of Figure \ref{fig: massinfluxandmassoutflux}, the blue solid curve shows  the mass influx at $x=50m$, the red solid curve shows the mass reflection at $x=50m$ and the green dotted curve is mass outflux at $x=150m$. 
As seen in Figure \ref{fig: massinfluxandmassoutflux}, the mass reflection has negative values.

In Table \ref{tab: massreflection}, the results for various amplitudes of the solitary wave are displayed for  $\Delta h=0.3$ on a slope $1:35$. 
Here  $\Delta h$ is the height of the topography. For $\Delta h=0.3$, the mass influx through the initial boundary of control volume is  defined by  
``$\text{Mass influx}=\int_{0}^{15}q_m(50,t) \,dt$", the mass outflux through the final boundary of control volume is  defined by  
``$\text{Mass outflux}=\int_{15}^{60}q_m(150,t) \,dt$" and the mass reflection through the initial boundary of control volume is  defined by  
``$\text{Mass reflection}=\int_{15}^{60}q_m(50,t) \,dt$". Note that the time limit may vary for other  $\Delta h$'s. The  error is defined by `` $ \text{error}= {\text{mass outflux - mass reflection - mass influx}}
$". It is clear from Table \ref{tab: massreflection}, that error  tends to $0$ as $\alpha=a / h_0$ approaches 0.

In Table \ref{tab: massreflection1}, the results for various $\Delta h$ of  water level are displayed with an initial amplitude $a=0.3$ on a slope $1:35$.
It is clear from Table \ref{tab: massreflection} and Table \ref{tab: massreflection1}, that the mass conservation holds approximately  for the coupled BBM system and the ratio between mass reflection and mass  influx is called ``mass ratio", 
which is smaller for smaller $\Delta h$.

\begin{table}[H]\footnotesize
  \begin{center}
    \begin{tabular}{| r | r | r | r | r|}
    \hline
   Amplitude&  Mass influx & Mass outflux & Mass reflection  & Error \\ \hline
    0.2&1.0995& 1.0117 & -0.0879 &  0.0001  \\ \hline
 0.3& 1.3856& 1.2799&-0.1059  &0.0002 \\ \hline
 0.4& 1.6438&  1.5236   & -0.1204 & 0.0002   \\ \hline
 0.5& 1.8856& 1.7529& -0.1329  & 0.0002 \\ \hline
 0.6&2.1166& 1.9732& -0.1438 & 0.0003 \\\hline
  \end{tabular}
\end{center}
 \caption{Error in mass conservation for different waveheights on a slope  1:35 and $\Delta h= 0.3$. 
  The ``$ \text{error}= {\text{mass outflux - mass reflection - mass influx}}$" quantifies the error in the mass balance law. This table suggest that mass conservation holds approximately. }
\label{tab: massreflection}
\end{table}

\begin{table}[H]\footnotesize
  \begin{center}
    \begin{tabular}{| r | r | r | r | r|r|}
    \hline
  $\Delta h$&  Mass influx & Mass outflux& Mass reflection  & Mass ratio
 \\  \hline
 0.1&1.3856&1.3535  &-0.0320 & 0.0232 \\ \hline
 0.2&1.3856& 1.3186 & -0.0669  &0.0483\\ \hline
 0.3&1.3856& 1.2799 &-0.1059   & 0.0764\\ \hline
 0.4&1.3856& 1.2358 &-0.1500   &0.1083\\ \hline
 0.6&1.3856& 1.1855&-0.2003 &0.1446\\  \hline
 
 \end{tabular}
\end{center}
\caption{The ratio between mass reflection and mass influx of a solitary wave with initial amplitude $a=0.3$ on a  
slope 1:35 for different $\Delta h$. Ratio of mass reflection and mass influx is decreasing with decreasing $\Delta h$.}
\label{tab: massreflection1}
\end{table}

The reflection of a small amplitude wave when a solitary wave goes through a slope is defined as ``reflection". To find  the ratio between  reflection and initial solitary wave,  the following  $L^2$-norm 
\begin{equation*}
\left \|\eta \right \|_{L^2(\mathbb{R})}^2=\int_{\mathbb{R}}\left | \eta(x) \right |^2 \,dx,
\end{equation*}
is used.

\begin{figure}[H]
 \centering
 \includegraphics[height=7cm]{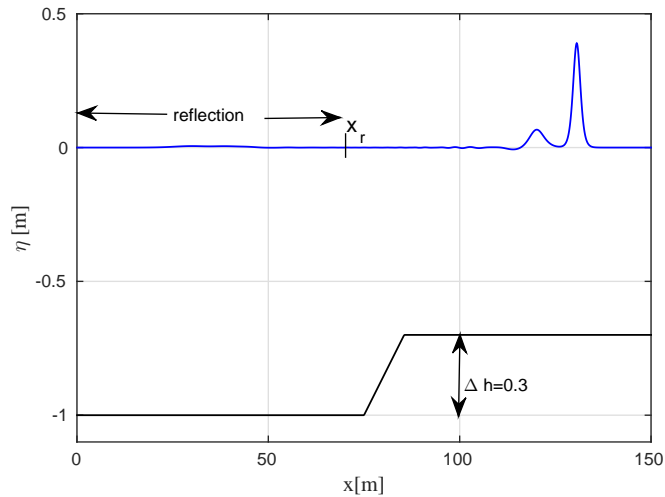}
  \vskip-0.1in
  \caption{ Reflection of solitary waves of initial amplitude $0.2m$ transformation on the slope $S=1:35$ and $\Delta h= 0.2$ in the physical domain.}
\label{fig:reflection}
\end{figure}

To calculate the  $L^2$-norm  of initial solitary waves, the value of $\eta$ is integrated with respect to $x$ on the fluid 
domain $[0,L]$ at initial time $t=0$. To determine the $L^2$-norm  of reflected wave, we  run the solitary wave on the slope for long  enough time to separate the reflection of small wave from  solitary wave.
The end point  of the reflected waves on the $x-$axis  is denoted by $x_r$ (see Figure \ref{fig:reflection}). Then  the reflected  wave is integrated on the interval $[0,x_r]$.
And the  corresponding ``reflection coefficient'' is then defined to be
\begin{equation*}
\frac{\left \|\eta _\text{ reflection}\right \|_{L^2([0,x_r])}}{\left \|\eta_ {\text{ initial at }  t=0 }\right \|_{L^2([0,L])}}.
\end{equation*} 
 It is clear from Table \ref{tab:reflection table} that the ``reflection coefficient'' approaches zero as the slope become more and more gentle. 
 \begin{table}[H]\footnotesize
\begin{center}
\begin{tabular}{lllllll}
\hline
\multicolumn{1}{|r|}{\multirow{2}{*}{ slope}} & \multicolumn{2}{l|}{$\Delta h=0.3, \ H_0=0.3$}                                                          & \multicolumn{2}{l|}{$\Delta h=0.2, \ H_0=0.2$}  &       \multicolumn{2}{l|}{$\Delta h=0.1, \ H_0=0.1$}                                                   \\ \cline{2-7} 
\multicolumn{1}{|r|}{}                      & \multicolumn{1}{l|}{$ \left \|\eta _\text{ reflection}\right \|_{L^2}$}   & \multicolumn{1}{l|}{Reflect. coeff.} & \multicolumn{1}{l|}{$ \left \|\eta _\text{ reflection}\right \|_{L^2}$}   & \multicolumn{1}{l|}{Reflect. coeff.} & \multicolumn{1}{l|}{$ \left \|\eta _\text{ reflection}\right \|_{L^2}$}   & \multicolumn{1}{l|}{Reflect. coeff.}\\ \hline
\multicolumn{1}{|l|}{1:35}                   & \multicolumn{1}{l|}{4.70e-04}  & \multicolumn{1}{l|}{1.70e-03}   & \multicolumn{1}{l|}{1.76e-04}& \multicolumn{1}{l|}{1.20e-03}       & \multicolumn{1}{l|}{2.61e-05}  & \multicolumn{1}{l|}{5.19e-04}  \\ \hline
\multicolumn{1}{|l|}{1:100}                   & \multicolumn{1}{l|}{1.77e-04}   & \multicolumn{1}{l|}{6.40e-04}   & \multicolumn{1}{l|}{7.11e-05}   & \multicolumn{1}{l|}{4.85e-04}        & \multicolumn{1}{l|}{1.45e-05}  & \multicolumn{1}{l|}{2.88e-04} \\ \hline
\multicolumn{1}{|l|}{1:400}                   & \multicolumn{1}{l|}{2.82e-05}  & \multicolumn{1}{l|}{1.02e-04}   & \multicolumn{1}{l|}{1.42e-05}  & \multicolumn{1}{l|}{9.69e-05}        & \multicolumn{1}{l|}{4.24e-06}  & \multicolumn{1}{l|}{8.43e-05} \\ \hline
\multicolumn{1}{|l|}{1:800}                   & \multicolumn{1}{l|}{8.73e-05}   & \multicolumn{1}{l|}{3.15e-05}   & \multicolumn{1}{l|}{3.04e-06}        & \multicolumn{1}{l|}{2.08e-05}      & \multicolumn{1}{l|}{3.56e-07}  & \multicolumn{1}{l|}{7.08e-06}   \\ \hline
                                                                                                 &                              &                            &                             &                          &                               &   
\end{tabular}
\caption{Calculation of the amount of ``reflected" waves for different slopes and amplitudes. It shows that  the ``reflection coefficient"  approaches  zero, as 
the slope becomes more and more gentle.}
\label{tab:reflection table}
\end{center}
\end{table}


\subsection{Evolution of solitary waves on a sloping  bottom }
\label{subsec:shoaling}

\begin{figure}[H]
  \centering
   
{
\includegraphics[height=7cm]{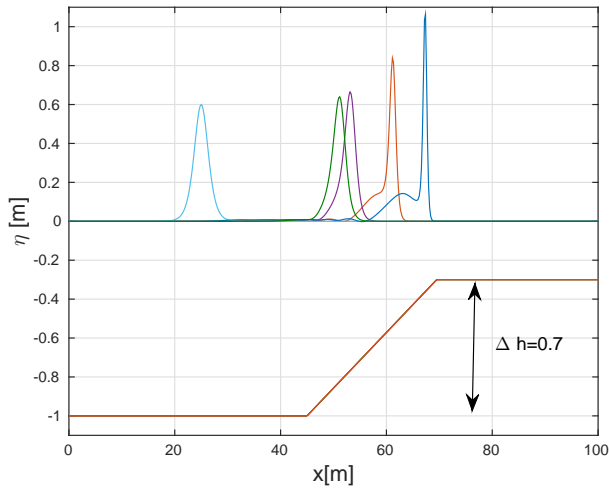}
}
\caption{Solitary wave of initial amplitude $0.6m$ transformation on the slope $S=1:35$. 
  Note that the bottom topography is  smoothed near the corners.}
   \label{fig: solitaryuneven}
\end{figure}

\begin{figure}[H]
  \centering
 \includegraphics[height=8cm]{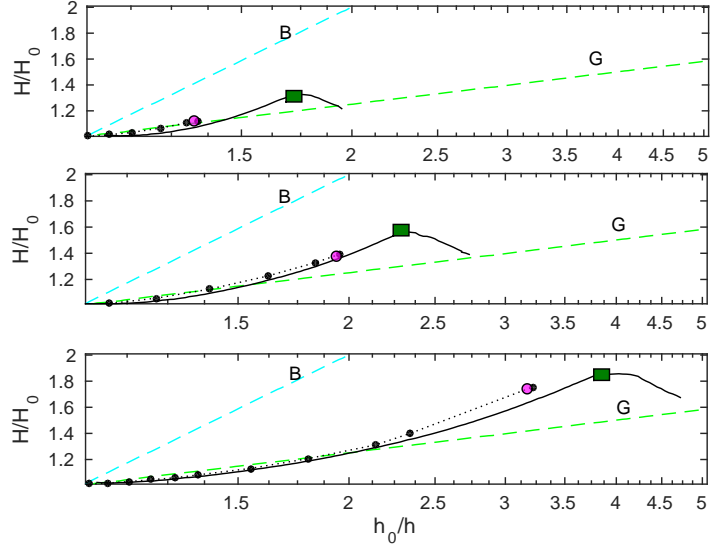}
  \vskip-0.1in
  \caption{Computations for the shoaling curves with initial amplitudes = 0.6 (upper panel);  0.4 (middle panel);  0.2 (lower panel) on a slope 1:35. Here G denotes Green's law, B denotes Boussinesq's law, the dotted curves are our numerical results  and the solid curve are
numerical results from  Grilli et al. \cite{GSS1997} .  Rectangular and circular symbols denote the breaking points of Grilli et al. \cite{GSS1997}  and present work respectively.}
  \label{fig: solitaryshoaling}
\end{figure}

 Shoaling of solitary waves  with different waveheights for initial undisturbed depth $h_0=1 m$ to smaller new depth up to $h=0.1 m$ are considered. 
The maximum waveheights were computed at different locations over the  slope $S=1:35$. Figure  \ref{fig: solitaryuneven} shows results  for solitary wave of height $0.6m$. It shows that
 waves crests become  steeper  while shoaling  on the slope. We generally see the reflection of a small amplitude wave when a solitary wave goes through a slope. After carefully 
 measuring waveheights over the different slopes the relative maximum local 
 waveheight $H/H_0$ versus the relative local depth $h_0/h$  are plotted in Figure \ref{fig: solitaryshoaling}, where $h,h_0,H$ and $H_0$ represent the local water depth, the constant reference water depth, 
 local solitary waveheight and initial solitary waveheight, respectively. For later reference, we define the shoaling rate to be the exponent $\alpha$ if the relation  $\frac{H}{H_0}=\left ( \frac{h_0}{h} \right )^{\alpha }$ holds.
  
 The effect of a varying bottom on water waves of this class is of obvious engineering importance and numerical solutions have been obtained by 
 Peregrine \cite{P1967} and  Madsen and Mei \cite{MM1969}  using a finite difference scheme to compute the deformation of a solitary wave climbing a beach. 
 Experimental results for wave shoaling and breaking of solitary waves were obtained by   Ippen and Kulin \cite{IK1954}, Kishi and Saeki \cite{KS1966}, Camfield and Street \cite{CS1969}  and Synolakis \cite{ST1987}. 
  Note also that Pelinovsky and Talipova \cite{PT1, PT2} studied the shoaling curves which were obtained by the waveheight–wave energy relation 
 for numerical solutions of the full water  wave problem found by Longuet-Higgins \cite{LH1974}, Longuet-Higgins and Fenton \cite{LF1974}. In case of a  periodic sequence of solitary waves, 
 Ostrovsky and Pelinovsky \cite{OP1} found that  the shoaling relation  reduces to a ''nonlinear'' Green's law.  
 Recently   the experimental results of Grilli et al. \cite{GSS1994} and numerical studies   based on potential flow theory for the Euler equations, which was presented by Grilli et al. \cite{GSS1997} have concentrated on shoaling studies.
 Noteworthy is the fact that the studies of  Grilli et al. \cite{GSS1994, GSS1997}  gives a nice picture of different shoaling regimes and predict a variety of scaling relations for the local wave amplitude
 ahead and beyond the breaking point.

 For comparison, we have considered the Grilli et al. \cite{GSS1997} numerical results.
 Figure \ref{fig: solitaryshoaling}  shows plots of data taken from \cite{GSS1997}.
 The shoaling curve for initial amplitudes 0.6 , 0.4  and 0.2 are plotted.  The Green's law, which predicts shoaling rates (amplitude increase) 
 $\propto h^{-1/4} $ is plotted with 'G' mark and the Boussinesq's law which gives shoaling rates  $\propto h^{-1}$ is  plotted with 'B' mark. 
  Figure \ref{fig: solitaryshoaling}  shows the  shoaling curves of the current work are in good agreement with the numerical results of Grilli et al. \cite{GSS1997}.
  It can be seen that the shoaling rate  increases initially more slowly than
predicted by Green’s law, but then increases as the water depth keeps decreasing.
 Although there is no breaking point in our numerical calculation, it is noticed that the breaking points appeared in the results obtained by Grilli et al. \cite{GSS1997}. 
 For instance, (\ref{eq:bvelocities}) is used to check the breaking criterion as discussed above.

  \begin{figure}[H]
\centering
\includegraphics[height=7cm]{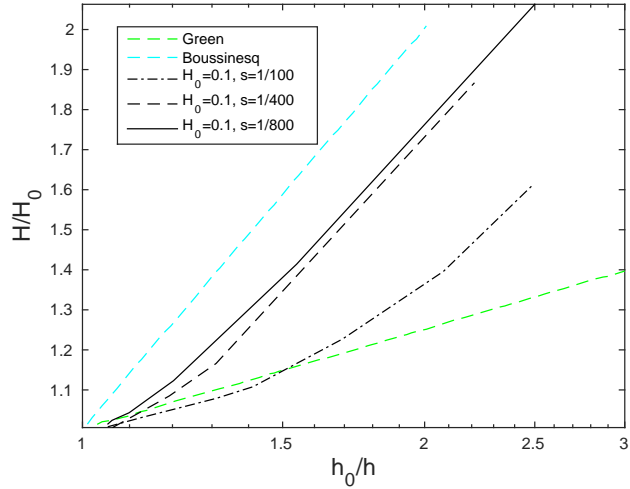}
\caption {Computations for the shoaling curves with initial amplitude  0.1 on different slopes $1:100$, $1:400$ and $1:800$. Here the solid curve represents the shoaling curve for the slope
$1:800$, the  dashed curve represents the shoaling curve for the slope $1:400$ and the dotted curve represents the shoaling curve for the slope $1:100$. }
\label{fig: comparison1}
\end{figure}

  Figure  \ref{fig: comparison1} shows plots of shoaling rates for wave of initial waveheight 0.1  with different slopes $1:100$, $1:400$ and $1:800$.  It is apparent that for slope $1:100$, the shoaling rate is lower than Green's law for small  $h_0/h$
  and higher for large  $h_0/h$.
  But for the smaller slopes $1:400$ and $1:800$, the shoaling rate is closer to the line  $h^{-1}$ for large  $h_0/h$. Apparently, the computed curves get close to Boussinesq's law for smaller slopes.

 Now  the breaking criterion (\ref{eq:bvelocities}) is applied to the solitary wave solutions. In order to find  wave breaking in these solitary wave solutions,  the $x$-location of the maximum waveheight at each time step is found.
 The propagation speed U is then estimated using these $x$-locations at each time step. Finally if the computed horizontal velocity  u exceeds the mean propagation speed U, we can conclude that, around this time step, the wave is starting to break.
  The water depth at breaking  measured under the wave crest is denoted  as $h_b$ and the corresponding solitary waveheight at 
 breaking is denoted as $H_b$. In Table \ref{tab: wavebreak table}   the relative breaking waveheight $H_b/h_b$ at the corresponding breaking points is compared with those of Grilli et al. \cite{GSS1997}
 and Chou et al. \cite{CSY2003}. 
 For a large wave amplitude the waveheight will exceed the breaking criterion very soon after it propagates on the slope and so wave breaking occurs almost instantly without too much change in height.
 The ratio of relative breaking waveheight is larger for small amplitude waves than for large amplitude waves. Wave breaking occurs sooner for larger initial waves. 
 
McCowan  \cite{MC1894} theoretically defined the breaker depth index as $H_b/h_b = 0.78$ for a solitary wave traveling over a horizontal bottom using the assumption
that instability is reached when the particle velocity at the crest equals the wave celerity and that the crest angle is then 120$^{\circ}$. 
To estimate the initial breaking waveheight on a mild-slope beach, this value ($H_b/h_b = 0.78$) is  most commonly used in engineering practice as a first estimate.
Ippen and Kulin \cite{IK1954}  showed that the upper limit of the breaking criterion should be 
0.78 for solitary wave over  very mild slope. In this article  the slope 1:35 is used
and it can be seen from Table \ref{tab: wavebreak table} that the relative breaking waveheights $H_b/h_b$  are smaller for higher amplitude waves. 
It is noticed that the  relative breaking waveheights $H_b/h_b$ at breaking points
are well above the McCowan limit  0.78. Since the  relative breaking waveheights $H_b/h_b$ at breaking points
are smaller than those obtained by Grilli et al. \cite{GSS1997}  and Chou et al. \cite{CSY2003}, we might consider higher order Boussinesq model for further study.

\begin{table}[H]\footnotesize
\begin{center}
\begin{tabular}{lllllll}
\hline
\multicolumn{1}{|r|}{\multirow{2}{*}{$H_0$}} & \multicolumn{3}{l|}{$H_b/h_b$}                                                          & \multicolumn{3}{l|}{$H_b/h_0$}                                                          \\ \cline{2-7} 
\multicolumn{1}{|r|}{}                      & \multicolumn{1}{l|}{Chou}  & \multicolumn{1}{l|}{Grilli} & \multicolumn{1}{l|}{Present} & \multicolumn{1}{l|}{Chou}  & \multicolumn{1}{l|}{Grilli} & \multicolumn{1}{l|}{Present} \\ \hline
\multicolumn{1}{|l|}{0.2}                   & \multicolumn{1}{l|}{1.330} & \multicolumn{1}{l|}{1.402}  & \multicolumn{1}{l|}{1.132}   & \multicolumn{1}{l|}{0.402} & \multicolumn{1}{l|}{0.364}  & \multicolumn{1}{l|}{0.3513}        \\ \hline
\multicolumn{1}{|l|}{0.25}                   & \multicolumn{1}{l|}{1.314} & \multicolumn{1}{l|}{1.385}  & \multicolumn{1}{l|}{1.056}   & \multicolumn{1}{l|}{0.465} & \multicolumn{1}{l|}{0.422}  & \multicolumn{1}{l|}{0.3984}        \\ \hline
\multicolumn{1}{|l|}{0.3}                   & \multicolumn{1}{l|}{1.283} & \multicolumn{1}{l|}{1.380}  & \multicolumn{1}{l|}{1.033}   & \multicolumn{1}{l|}{0.514} & \multicolumn{1}{l|}{0.476}  & \multicolumn{1}{l|}{0.4475}        \\ \hline
\multicolumn{1}{|l|}{0.4}                   & \multicolumn{1}{l|}{1.26}  & \multicolumn{1}{l|}{1.378}  & \multicolumn{1}{l|}{0.977}   & \multicolumn{1}{l|}{0.614} & \multicolumn{1}{l|}{0.592}       & \multicolumn{1}{l|}{0.5320}        \\ \hline
                                            &                            &                             &                              &                            &                             &                             
\end{tabular}
\caption{Comparison of the relative breaking waveheight for waves with initial amplitudes 0.2, 0.25, 0.3, 0.4 on slope 1:35.}
\label{tab: wavebreak table}
\end{center}
\end{table}

 \subsection{Comparison of mild slope model systems and results}
 \label{subsec:model equations}
 
\begin{figure}[H]
\centering
\includegraphics[height=7cm]{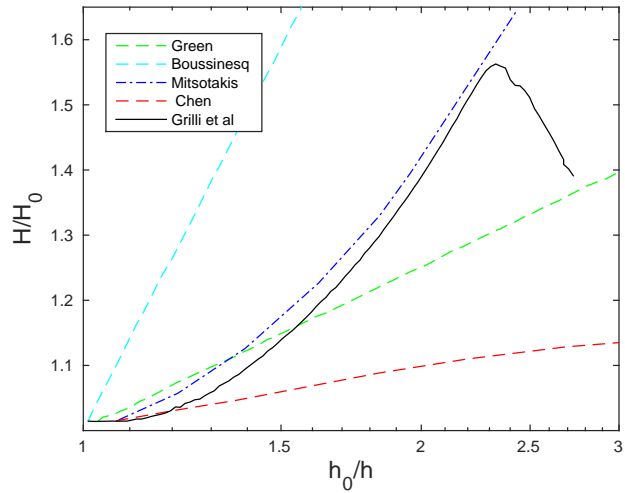}
\caption{Computations for the shoaling curves with initial amplitude 0.4  on a slope 1:35. Here the dashed - dotted curve represents  numerical results for the system (\ref{eq:BBmsystem}) derived by Mitsotakis \cite{MDE2009}, the solid curve are
numerical results from Grilli et al. \cite{GSS1997} and the dashed curve represents numerical results for the system (\ref{eq:chenBBmsystem}). Indeed the system (\ref{eq:chenBBmsystem}) works for small-amplitude bottom variations 
as expected, since the bottom function $h(x)$ is assumed to be of order  $\mathcal{O}(\alpha)$.}
\label{fig: comparison2}
\end{figure}

For comparison,  the work of Chen \cite{C2003} is considered. Chen presented equations for bi-directional waves over an uneven bottom,
which may be written in non-dimensional,  unscaled variables and disregard terms of order 
$\mathcal{O}(\alpha \beta,\beta ^2)$ as

\begin{subequations} \label{eq:chenBBmsystem}
 \begin{eqnarray}
u_{t}+g {\eta }_{x}+ {u}{u}_{x}
 -\frac{1}{2}\left ( 1- \theta ^2\right ) {h_0}^2 u_{xxt}  =0, \\
{\eta }_{t}+\left ( {{\eta }}u+{h}{u} \right )_{x}
 - \frac{1}{2}\left (  \theta ^2-\frac{1}{3}\right ){h_0}^2 {\eta}_{xxt} =0.
 \end{eqnarray}
\end{subequations}
The Chen and Mitsotakis models \cite{C2003, MDE2009}  represent the same type of coupled BBM-type system,  derived in the context of the Boussinesq scaling. One can derive a number of special cases of the general Boussinesq system. 
Since we are interested in coupled BBM-type system,  the model of Chen is chosen for comparisons.
The above system (\ref{eq:chenBBmsystem}) is solved using the same numerical technique above. The main difference between the two systems (\ref{eq:BBmsystem}) and (\ref{eq:chenBBmsystem}) is
approximation of bottom motion. In (\ref{eq:BBmsystem}), the bottom motion is nondimensionalized by $\tilde{h}=\frac{h}{h_0}$, and in (\ref{eq:chenBBmsystem}), it is nondimensionalized by 
$\tilde{h}=\frac{h-h_0}{a_0}$ which is similar to the approximation of  wave amplitude $\eta$.  Figure \ref{fig: comparison2} shows computations for the shoaling curves with initial amplitude $0.4m$.  
It is noticed that  the shoaling curve corresponding to the system (\ref{eq:chenBBmsystem}) lies below the  Green’s law because of the lower order approximation. 
In the paper  \cite{C2003}, the bottom function $h(x)$ is assumed to be   $\mathcal{O}(\alpha)$ and  in the paper  \cite{MDE2009}, the bottom function $h(x)$ is assumed to be   $\mathcal{O}(1)$.
The results are in line with the assumptions used in their respective derivations.

\section{Conclusion}
\label{sec:conclusion}
In this article, a coupled BBM system of equations has been studied in the situation of   water waves propagating  over decreasing fluid depth. A conservation equation for mass and a wave breaking 
criteria, both valid in the Boussinesq approximation have been found. 
A Fourier collocation  method  coupled with a 4-stage Runge-Kutta time integration scheme has been employed in this work to approximate the solution of the BBM system.
It has been shown that the approximate mass conservation relation is  reasonably accurate.
Moreover, the results from evaluation of the approximate mass conservation law show that  the ratio of mass reflection to mass influx approaches  zero as $\Delta h$ becomes small.

In our previous paper \cite{KS2013} we showed that for waves of very small amplitude, the shoaling relation approaches Boussinesq's law for Boussinesq-type systems which are valid for waves with the Stokes number 
$S= \alpha/\beta$  of order 1, and in this case we measured the transition of the wave  only at the initial and final stage assuming the wave undergoes an adiabatic  adjustment. 
It is confirmed from Table  \ref{tab:reflection table} that the $L^2-$ratio between  reflection and initial solitary wave  approaches  zero as  the slope becomes more and more gentle which lends additional credibility to
shoaling results based on adiabatic approximation. In addition the results  displayed in Figure \ref{fig: comparison1} indicate that shoaling rates for small amplitude waves are closer to Boussinesq's 
law for very gentle  slopes.

Considering shoaling of finite amplitude waves, we have compared shoaling curves obtained with the current method to numerical results of Grilli et al. \cite{GSS1997}  for the Euler equations based on potential flow theory, and
the corresponding shoaling curve of the current work is in good agreement with the numerical results of Grilli et al. \cite{GSS1997} and it has been found that the variation in waveheight of a shoaling 
solitary wave initially increases at a lower rate than that predicted by Green’s law, but  then increases similar to Boussinesq’s
law. Indeed, the shoaling curves  achieved in this paper matches the shoaling curves of  Grilli et al. \cite{GSS1997}
better than the similar approximation established by Khorsand and Kalisch \cite{KK2014}. 

The comparison of shoaling curves of two model systems (\ref{eq:BBmsystem}) (\cite{MDE2009}) and (\ref{eq:chenBBmsystem}) (\cite{C2003}) with the numerical results of Grilli et al. \cite{GSS1997}
showed that  each of these models works well in their respective regimes of applicability.

\section*{Acknowledgement}

I would like  to thank  my supervisor, Professor Henrik Kalisch for his helpful guidance, comments and detailed correction on this work. This research was supported by Research council of Norway.

\end{document}